\documentclass[cits]{PoS}
\usepackage{graphicx}
\usepackage{rotating}

\title{Evidence of powerful relativistic jets in narrow-line Seyfert 1 galaxies}

\ShortTitle{Relativistic jets in NLS1}

\author{\speaker{Luigi Foschini}%
         \thanks{I would like to thank for the interesting, useful, and amusing discussions G. Ghisellini, L. Maraschi, and F. Tavecchio. Thanks also to S. Komossa for the editorial handling of this manuscript.}\\
        INAF Osservatorio Astronomico di Brera, Via E. Bianchi 46, 23807, Merate (Italy)\\
        E-mail: \email{luigi.foschini@brera.inaf.it}}

\abstract{In 2008, the \emph{Fermi Gamma-ray Space Telescope} has revealed -- for the first time -- high-energy (E~$>$100~MeV) $\gamma$ rays from a few Narrow-Line Seyfert 1 Galaxies (NLS1s). Later, in 2009 and 2010, two multifrequency campaigns on one of these sources, PMN~J0948$+$0022 ($z=0.585$), definitely confirmed the presence, in sources of this type, of a relativistic jet very similar and with comparable power to those in blazars. However, these sources are neither blazars nor radio galaxies, as proven by their optical spectrum and by their very compact radio morphology. Moreover, since NLS1s are generally hosted in spiral galaxies, this casts a significant doubt on the paradigm of the correlation between jets and elliptical host galaxies. These findings pose intriguing challenges to the current knowledge of jet systems and on how these structures are generated. The current status of the researches in this field is reviewed and ongoing work is outlined.}

\FullConference{Narrow-Line Seyfert 1 Galaxies and their place in the Universe\\
		 April 4-6, 2011\\
		 Milano, Italy}

\begin{document}

\section{Introduction}
Narrow-Line Seyfert 1 Galaxies (NLS1s) constitute a class of active galactic nuclei (AGNs) with peculiar characteristics. Before this workshop, I was aware that NLS1s were very interesting sources, but now, after the conclusion of the meeting and having heard several interesting talks, I have understood that NLS1s are even more interesting and special than I could have thought (\footnote{See the workshop proceedings at \texttt{http://pos.sissa.it/cgi-bin/reader/conf.cgi?confid=126}.}). 

Richard Pogge in his talk \cite{POGGE} has outlined 25 years of NLS1s studies, from the first interesting notes (e.g. \cite{DAVIDSON,GASKELL}), to the awareness of the discovery of a new class of AGNs \cite{OP}. Several authors have pointed out that the observed features at different wavelengths suggest that NLS1s are AGNs with relatively low masses, high accretion rates, and possibly young (e.g. \cite{BOLLER,BOROSON1,BOROSON2,GRUPE1,MATHUR,PETERSON}; see the updates \cite{BOROSON3,GRUPE3,MATHUR3,PETERSON2}). NLS1s generally do not display strong activity at radio frequencies, but a small percent of them ($\sim 7$\%) do not follow this rule \cite{KOMOSSA}.  

The first radio-loud NLS1, PKS~0558$-$504 (z=0.137), was discovered in 1986 \cite{REMILLARD}. Later, in the early 2000s, a few more sources of this type were discovered \cite{GRUPE2,OSHLACK,ZHOU1,KOMOSSA} and there was already a first (unsuccessful) attempt to detect very high-energy $\gamma$ rays ($E>400$~GeV) from these sources with the \emph{Whipple} \v{C}erenkov telescope. More radio-loud NLS1s were found with early surveys both radio- and optically-selected \cite{WHALEN,YUAN}. Although, in some cases, the flat or inverted radio spectrum and high brightness temperature suggested the presence of a relativistic jet \cite{ZHOU1,DOI}, in other cases there was no direct indication of beaming \cite{KOMOSSA}. Moreover, these sources had a very compact radio morphology, making it difficult to trace the components searching for superluminal motion \cite{DOI}. Other indications of a possible presence of a relativistic jet came from optical-to-X-ray studies of variability \cite{FOS1}.

The definitive proof of the existence of relativistic jets in NLS1s arrived after the launch of the \emph{Fermi Gamma-ray Space Telescope} (hereafter \emph{Fermi}) in 2008 June. Already after a few months of operation, there was the first detection at $\gamma$ rays ($E>100$~MeV) of a NLS1: it was PMN~J0948$+$0022 ($z=0.5846$) \cite{ABDO1,FOS2}. This discovery immediately triggered a multiwavelength (MW) campaign (2009 March-July), which confirmed that the $\gamma$-ray emission was indeed associated with the NLS1 and that the MW behavior was typical of a relativistic jet, like those in blazars \cite{ABDO2,GIROLETTI}. The NLS1 had a peak in the $\gamma$-ray flux on April 1, 2009, with a value of $\sim 4\times 10^{-7}$~ph~cm$^{-2}$~s$^{-1}$ and then declined, followed by a similar trend at all the wavelengths, while radio emission increased reaching the peak about a couple of months later. In 2010 July, PMN~J0948$+$0022 underwent a much stronger outburst, reaching the power of $\sim 10^{48}$~erg/s in the $0.1-100$~GeV band and this event was preceded by a significant change in the radio polarization angle, which occurred about one year before \cite{FOS3}. It is worth noting the extreme power at $\gamma$ rays together with the lack of extended radio structures. One basic question, stressed by G. Ghisellini, is indeed: where such great power has gone?

This is just one of the many problems opened by the discovery of high-energy $\gamma$-ray emission from NLS1s, but many others are present. I am not able yet to say what is the effective impact of this discovery in our knowledge of AGNs and relativistic jets. I am still in the phase of searching and collecting informations to better understand the properties of these sources, compare with other known AGNs, and try to infer some useful knowledge. Although there is a handful of $\gamma$-NLS1s, in the present work I would like to report about an early study of their MW properties and the comparison with radio-quiet objects and with bright $\gamma$-ray blazars.

\begin{figure}[!t]
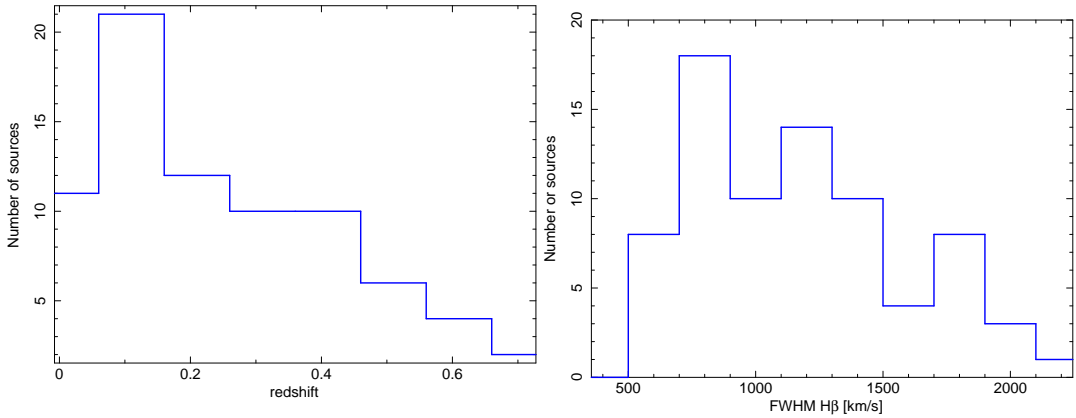

\begin{center}
\includegraphics[angle=270,scale=0.3]{histo_z.ps}
\includegraphics[angle=270,scale=0.3]{histo_FWHM.ps}
\vskip 24pt
\caption{Sample of NLS1s: (\emph{left panel}) Redshift distribution; (\emph{right panel}) FWHM H$\beta$ distribution.}
\label{fig:population}
\end{center}
\end{figure}

\section{Sample selection}
Rather obviously, I performed the early search for high-energy $\gamma$ rays from NLS1s in 2008 by starting with a sample built with the most radio-loud objects. The core of this sample was the list of very radio-loud sources of Yuan et al. \cite{YUAN}, which were optically selected from the \emph{Sloan Digital Sky Survey (SDSS)} and had the radio loudness $R=f_{1.4~\rm GHz}/f_{440~\rm nm} > 100$. Some sources with the same $R$ prescription were found in the literature and added to the sample \cite{GALLO,KOMOSSA,OSHLACK,ZHOU2,ZHOU3}. This list was composed of 29 very radio-loud NLS1s, which resulted in 4 detections at high-energy $\gamma$ rays after one year of \emph{Fermi} operations \cite{ABDO3}.

Now, in order to make a more extended work, I have relaxed the prescription on $R$ and added also less radio-loud and even radio-quiet (although not silent) NLS1s from the \emph{FIRST Bright Quasars Survey (FBQS)} \cite{WHALEN}. Therefore, the sample studied in the present work is composed of 76 NLS1s, of which 45 are from the FBQS \cite{WHALEN}, 23 are from the SDSS \cite{YUAN} and 8 are from literature \cite{GALLO,KOMOSSA,OSHLACK,ZHOU2,ZHOU3} (see Tables~\ref{table:sample} and \ref{table:obs} in the Appendix). 
The radio-quiet ($R<10$) subsample is composed of 30 sources, while the radio-loud one ($R>20$) has 39 objects; 7 NLS1s of the present list are radio-intermediate ($10<R<20$). The distributions of $z$ and FWHM H$\beta$ are displayed in Fig.~\ref{fig:population}. The measurements of the FWHM H$\beta$ are taken from the available literature (\cite{GALLO,KOMOSSA,OSHLACK,WHALEN,YUAN,ZHOU2,ZHOU3}).

I have also considered, as a comparison, a subsample of the bright $\gamma$-ray blazars detected by \emph{Fermi} during its early months of operations \cite{LBAS}. The subsample of 34 AGNs has been defined on the basis of the availability of \emph{Swift} optical ($B$ filter) and X-ray data \cite{GHISELLINI1,GHISELLINI2}. Radio data were taken from the \emph{Faint Images of the Radio Sky at Twenty Centimenters} (FIRST), as for the NLS1s (see next Section). It is worth noting that the original selection of these blazars was done on the basis of the high detection significance ($>10\sigma$, \cite{LBAS}), so this is not a flux-limited sample. On the other hand, also the sample of NLS1s is not flux limited.

\section{Multiwavelength Data}
For all the sources in Table~\ref{table:sample}, I gathered the data from radio to $\gamma$ rays from the publicly available archives and data servers (Table~\ref{table:obs}). All the data were corrected for the absorption due to the Galactic hydrogen, transformed -- when necessary -- into flux density at specific frequency (radio: 1.4~GHz; optical: 440~nm; X-rays: 1~keV; $\gamma$ rays: 100~MeV), \emph{K}-corrected by multiplying the flux density with the factor $(1+z)^{\alpha-1}$, having assumed the following values of $\alpha_{\rm r}=0$, $\alpha_{\rm o}=-0.5$ (see \cite{YUAN}), $\alpha_{\rm X}=1$, and $\alpha_{\rm \gamma}=1.5$ (see \cite{ABDO3}). The corresponding luminosities have been calculated within a $\Lambda$CDM cosmology with $H_{0}=70$~km~s$^{-1}$~Mpc$^{-1}$ and $\Omega_{\rm \Lambda}=0.73$ \cite{KOMATSU}. 

\subsection{Radio data}
Radio data were taken from the VLA (\emph{Very Large Array}) FIRST (\emph{Faint Images of the Radio Sky at Twenty Centimenters}\footnote{\texttt{http://sundog.stsci.edu/index.html}}) survey \cite{BECKER}, operated at 1.4~GHz. Only two sources were in sky regions not covered by the FIRST: PMN~J0134$-$4258 and PKS~0558$-$504. In both cases, I extended to 1.4~GHz the flux density measured at 4.8~GHz \cite{WRIGHT} by adopting a flat radio spectrum ($\alpha_{\rm r}=0$). 

\subsection{Optical data}
Most of the optical data in the $B$ filter band (440~nm) were taken from the \emph{Sloan Digital Sky Survey} (SDSS\footnote{\texttt{http://www.sdss.org/}}). The $ugriz$ system was converted into the $UBVRI$ one by using the formulae reported in \cite{CHONIS}. When the source was not in a region covered by the SDSS, I used the measurements of the \emph{US Naval Observatory} (USNO) B1 Catalog \cite{MONET}. For PKS~0558$-$504, I refer to \cite{OJHA}.

The observed magnitudes have been dereddened by using the $A_{\rm V}$ calculated from the $N_{\rm H}$ measurements of the \emph{Leiden-Argentine-Bonn Survey} \cite{KALBERLA} and the extinction laws by \cite{CARDELLI}.

Most of the sources are quasar-like and, therefore, the host galaxy contribution is negligible. For the lowest redshift sources, I estimated that the host galaxy contributed at maximum by a 10\% in flux, by assuming a spiral host of Sa type\footnote{A S0 type would be fainter. See \cite{VANDENBERGH}.}.

\subsection{X-ray data}
I have used the \emph{ROSAT Bright and Faint Source Catalogues}(\footnote{\texttt{http://www.xray.mpe.mpg.de/cgi-bin/rosat/rosat-survey}}) \cite{VOGES1,VOGES2}, which are based on an all-sky survey in the 0.1$-$2.4~keV energy band. When there is no detection, I calculated a $3\sigma$ upper limit by using the available exposure at the source position. In a few cases, although there was no detection with \emph{ROSAT}, the NLS1 was later detected with other satellites (e.g. PKS~1502$+$036 with \emph{Swift}, \cite{ABDO3}). However, for the sake of homogeneity, I have preferred to keep the upper limit. 

The count rates were then dereddened and converted into physical units (c.g.s.) by means of \texttt{WebPIMMS}(\footnote{\texttt{http://heasarc.gsfc.nasa.gov/Tools/w3pimms.html}}), having frozen the $N_{\rm H}$ to the Galactic value (\cite{KALBERLA}) and the photon index of the power-law model to $2$. 

\subsection{$\gamma$-ray data}
I have retrieved all the publicly available data of the Large Area Telescope (LAT, \cite{ATWOOD}) onboard \emph{Fermi} included in the period 2008 August 5 00:00 UTC to 2011 February 2 00:00 UTC, i.e. about 30 months of data (\footnote{\texttt{http://fermi.gsfc.nasa.gov/cgi-bin/ssc/LAT/LATDataQuery.cgi}}). The analysis was performed by using the \texttt{LAT Science Tools v. 9.18.6} and the corresponding background and instrument response files (\footnote{\texttt{http://fermi.gsfc.nasa.gov/ssc/data/analysis/}}). The adopted techniques are quite standard and are described with more details in other papers (e.g. \cite{ABDO1,ABDO2,ABDO3}). 

First, I have searched for new detections by integrating all the 30 months of data. To take into account the possible contaminating sources inside the 10$^{\circ}$ radius of interest centered on the NLS1 under examination, I started modeling the sources in the first \emph{Fermi} catalog \cite{ABDO4}. If, after an inspection of the results of the first analysis run, I found some new contaminating sources, then I added them to the list of modeled sources and repeated the run. The iteration stops when no more newly contaminating sources were found. In addition to the confirmation of the already known $\gamma$-NLS1s \cite{ABDO3}, I have found three new candidates. The results are summarized in Table~\ref{tab:latdetections}.

\begin{table}[!t]
\scriptsize
\begin{center}
\begin{tabular}{lrrcccccc}
\hline
Name & RA & Dec & err(dist) & $F_{0.1-100\rm~GeV}$ & $\Gamma$ & TS & $|\tau|$ & S/N\\ 
\hline
1H~0323$+$342 & 51.25 & $+$34.20 & 0.12(0.07) & 6.0$\pm$0.7 & 2.87$\pm$0.09 & 164 & $<$2.7 & 4.0\\
\emph{SBS~0846$+$513} & 132.45 & $+$51.19 & 0.11(0.05) & 0.51$\pm$0.15 & 2.0$\pm$0.1 & 52 & 12$\pm$8 & 4.7\\
PMN~J0948$+$0022 & 147.253 & $+$0.385 & 0.07(0.02) & 13.7$\pm$0.7 & 2.85$\pm$0.04 & 1081 & $<$0.8 & 5.4\\
\emph{FBQS~J1102$+$2239} & 165.70 & $+$22.63 & 0.37(0.10) & 2.0$\pm$0.6 & 3.1$\pm$0.2 & 32 & 25$\pm$12 & 2.9\\
\emph{SDSS~J124634.65$+$023809.0} & 191.83 & $+$2.53 & 0.47(0.21) & 1.7$\pm$0.7 & 3.1$\pm$0.3 & 15 & 32$\pm$15 & 2.1\\
PKS~1502$+$036 & 226.257 & $+$3.457 & 0.05(0.02) & 7.0$\pm$0.6 & 2.71$\pm$0.07 & 411 & 1.3$\pm$0.5 & 6.6\\
PKS~2004$-$447 & 302.002 & $-$44.504 & 0.08(0.07) & 1.2$\pm$0.3 & 2.3$\pm$0.1 & 44 & 6.2$\pm$1.7 & 12\\
\hline
\end{tabular}
\end{center}
\caption{Summary of \emph{Fermi}/LAT detections of $\gamma$-NLS1s based on 30 months of data. The new detections are emphasized in italic. The columns ``RA'' and ``Dec'' (J2000) indicate the position of the $\gamma$-ray source. The column ``err(dist)'' is the error at 95\% confidence level [deg] and, between parentheses, there is the distance [deg] of the $\gamma$-ray centroid from the radio coordinates. The three following columns indicate -- respectively -- the integrated $\gamma$-ray flux in the 0.1$-$100~GeV energy band [$10^{-8}$~ph~cm$^{-2}$~s$^{-1}$], the photon index of the power-law model used to fit the $\gamma$ photons, and the test statistic (TS, see \cite{MATTOX} for its definition), where $\sqrt{TS}\sim \sigma$. The two latest columns, $\tau$ and S/N, display the characteristic time scale [days] of doubling/halving flux and the significance of the flux change [$\sigma$], respectively.}
\label{tab:latdetections}
\normalsize
\end{table}

\begin{figure}[!h]
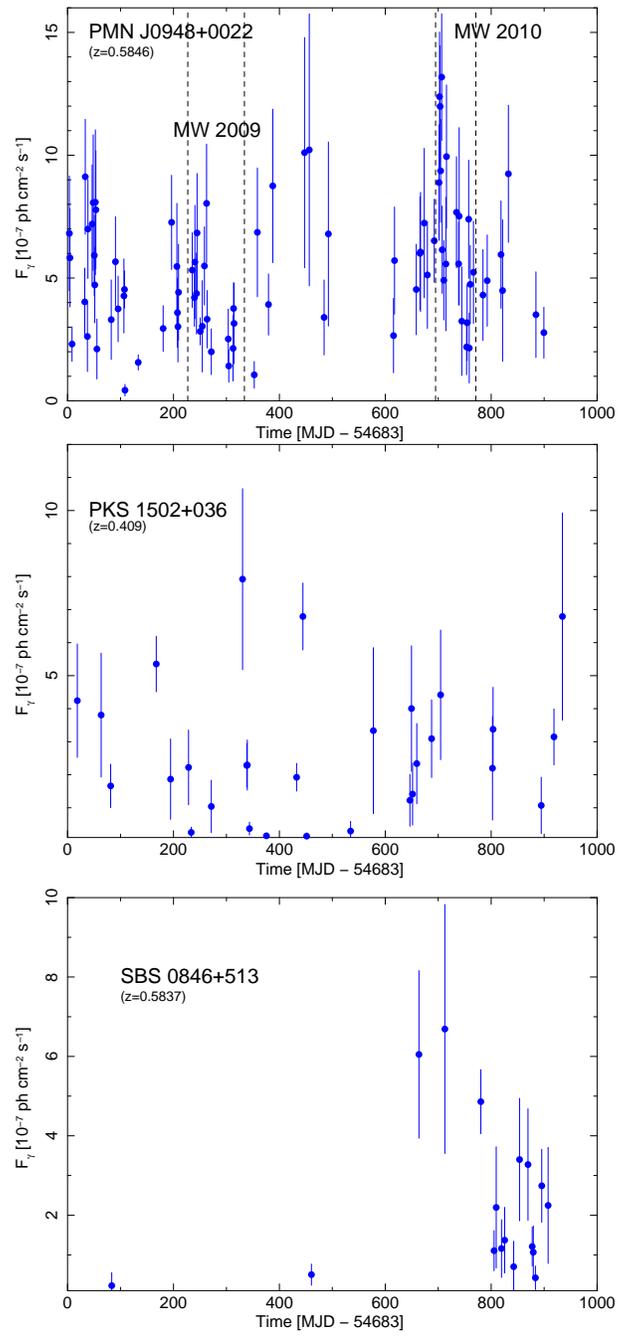

\begin{center}
\includegraphics[angle=270,scale=0.33]{0948_curva1d.ps}\\
\includegraphics[angle=270,scale=0.33]{1502_curva1d.ps}\\
\includegraphics[angle=270,scale=0.33]{0846_curva1d.ps}\\
\vskip 24pt
\caption{Examples of \emph{Fermi}/LAT light curves of NLS1s (0.1$-$100~GeV; 1 day time bin). (\emph{top panel}) PMN~J0948$+$0022 was the first NLS1 to be detected at $\gamma$ rays \cite{ABDO1, FOS2} and was also the target of two MW campaigns in 2009 \cite{ABDO2} and 2010 \cite{FOS3}, whose periods are indicated with dashed vertical lines; (\emph{center panel}) PKS~1502$+$036 was detected in the first year of operations \cite{ABDO3}; (\emph{bottom panel}) SBS~0846$+$513 is a newly detected $\gamma$-NLS1, because it becomes active during the latest months. }
\label{fig:curveLAT}
\end{center}
\end{figure}

I have also studied the variability at $\gamma$ rays of these sources, by producing light curves in the 0.1$-$100~GeV energy band with different binnings (1, 2, and 3 days) and taking as good the bins with detections at level $TS>10$ ($\sim 3\sigma$). The obtained curves consist of a number of bins spanning from a few points to 87 in the case of PMN~J0948$+$0022, the brightest $\gamma$-NLS1. A few examples of light curves are shown in Fig.~\ref{fig:curveLAT}. It is worth noting the case of SBS~0846$+$513, one of the three newly discovered $\gamma$-NLS1s reported in this work: the source was discovered since it became active during the latest months. As I will show also later, this means that the discovery of new $\gamma$-NLS1s ultimately depends on the status of activity of the source. 

\begin{figure}[!t]
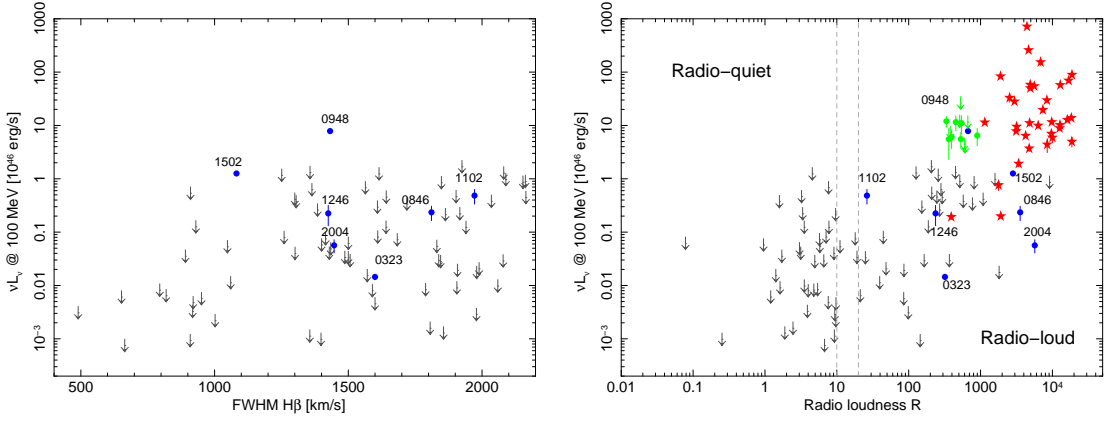

\begin{center}
\includegraphics[angle=270,scale=0.3]{lgamma_fwhm.ps}
\includegraphics[angle=270,scale=0.3]{lgamma_RL.ps}
\vskip 24pt
\caption{Sample of NLS1s: $\gamma$ rays vs FWHM H$\beta$ (\emph{left panel}) and radio loudness (\emph{right panel}). The detections with \emph{Fermi}/LAT are indicated with blue filled circles. The gray arrows indicate upper limits at $5\sigma$. The green points refer to the 2009 MW Campaign of PMN~J0948$+$0022 \cite{ABDO2}. The red stars are for the bright $\gamma$-ray blazars from the list of \cite{GHISELLINI1,GHISELLINI2}. The two dashed vertical lines represent the separation zone between radio-quiet and radio-loud sources. }
\label{fig:sample1}
\end{center}
\end{figure}

Then, I have analyzed all the lightcurves searching for flux variations greater than $3\sigma$ and calculated the corresponding time scale $\tau$ for a doubling/halving flux, defined as:

\begin{equation}
\frac{F(t)}{F(t_0)} = 2^{-\frac{(t-t_0)}{\tau}}
\end{equation}

The results are displayed in Table~\ref{tab:latdetections}. I have reported also the cases of FBQS~J1102$+$2239 and SDSS~J124634.65$+$023809.0, although the flux variations are not significant. These are also the two faintest $\gamma$-NLS1s, with the poorest statistics for the moment. A better analysis will hopefully be done in the future, if the sources will display some enhanced activity.

\begin{figure}[!t]
\begin{center}
\includegraphics[angle=270,scale=0.5]{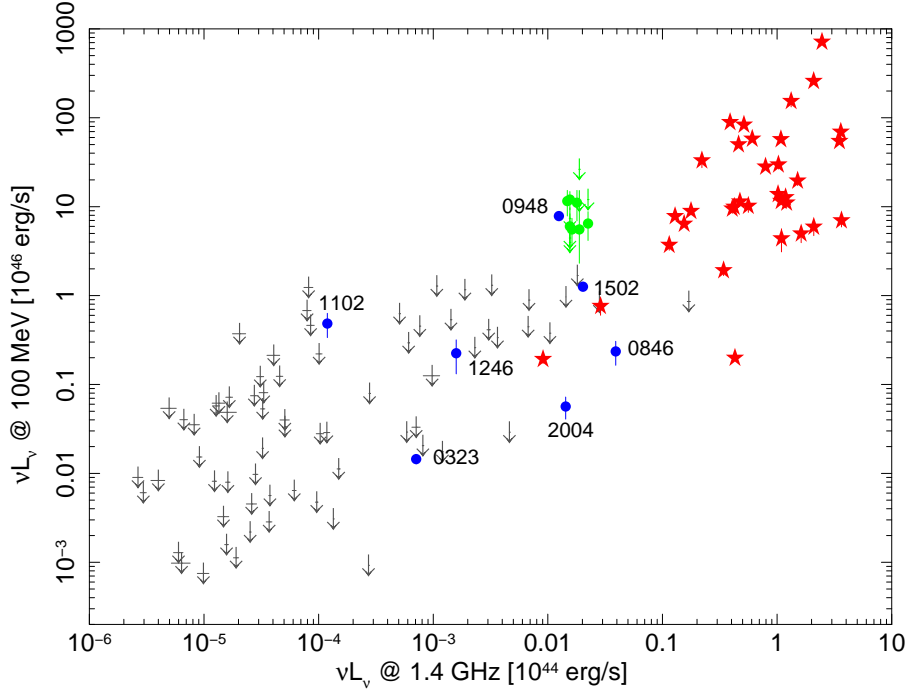}
\vskip 24pt
\caption{Sample of NLS1s: $\gamma$ rays vs radio. The detections with \emph{Fermi}/LAT are indicated with blue filled circles. The gray arrows indicate upper limits at $5\sigma$. The green points refer to the 2009 MW Campaign of PMN~J0948$+$0022 \cite{ABDO2}. The red stars are for the bright $\gamma$-ray blazars from the list of \cite{GHISELLINI1,GHISELLINI2}.}
\label{fig:radio}
\end{center}
\end{figure}

\begin{figure}[!t]
\begin{center}
\includegraphics[angle=270,scale=0.5]{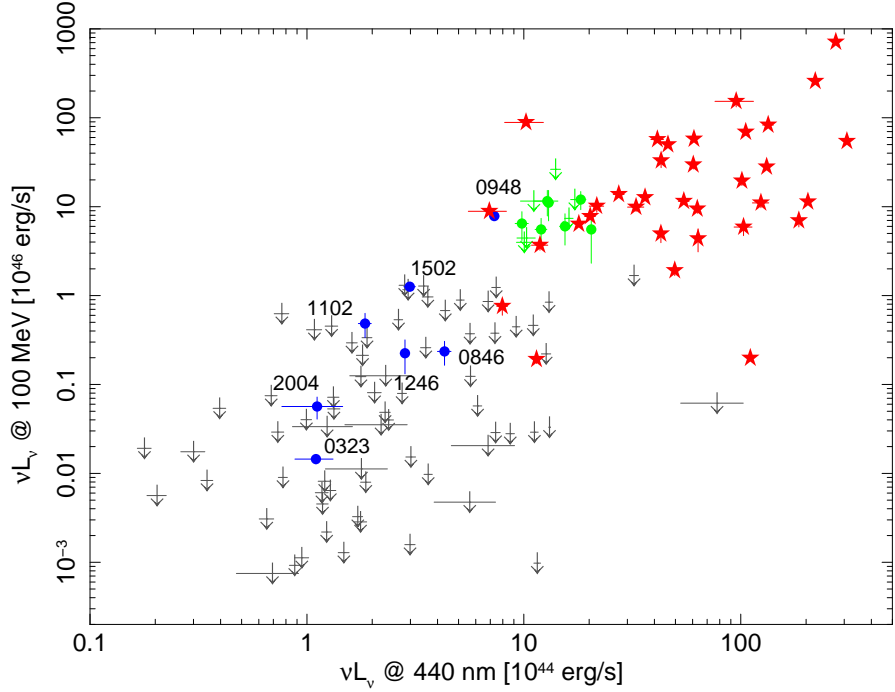}
\vskip 24pt
\caption{Sample of NLS1s: $\gamma$ rays vs optical. The detections with \emph{Fermi}/LAT are indicated with blue filled circles. The gray arrows indicate upper limits at $5\sigma$. The green points refer to the 2009 MW Campaign of PMN~J0948$+$0022 \cite{ABDO2}. The red stars are for the bright $\gamma$-ray blazars from the list of \cite{GHISELLINI1,GHISELLINI2}.}
\label{fig:optical}
\end{center}
\end{figure}

\begin{figure}[!t]
\begin{center}
\includegraphics[angle=270,scale=0.5]{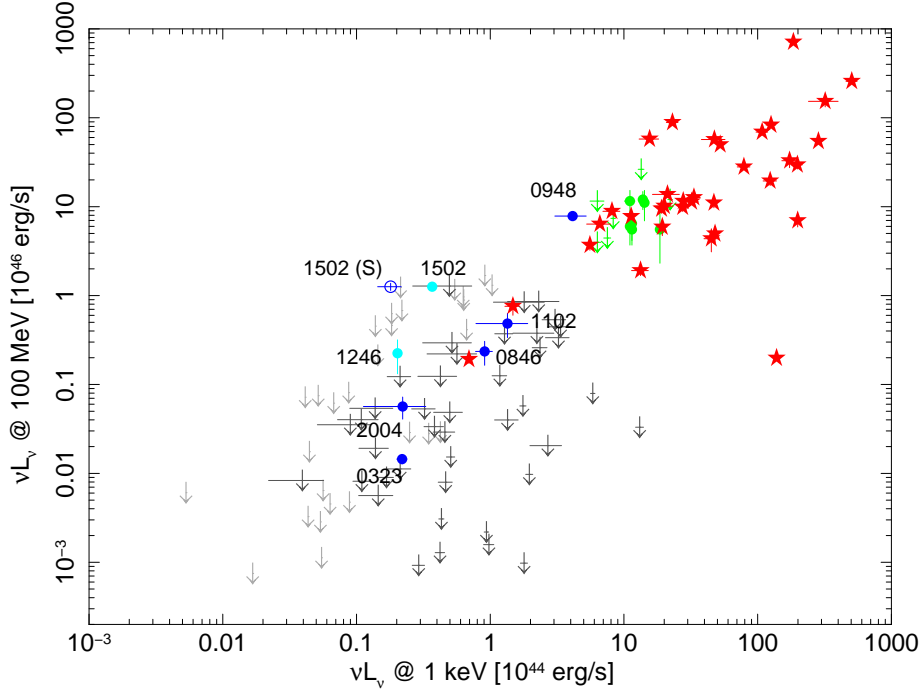}
\vskip 24pt
\caption{Sample of NLS1s: $\gamma$ rays vs X-rays. The detections with \emph{Fermi}/LAT are indicated with blue filled circles. The gray arrows indicate upper limits at $5\sigma$. The green points refer to the 2009 MW Campaign of PMN~J0948$+$0022 \cite{ABDO2}. The red stars are for the bright $\gamma$-ray blazars from the list of \cite{GHISELLINI1,GHISELLINI2}. If the source is not detected by \emph{ROSAT}, then it is reported the $3\sigma$ upper limit. The sources not detected both by \emph{ROSAT} and \emph{Fermi} are indicated with light gray arrows, while the sources detected by \emph{Fermi}, but not by \emph{ROSAT} are in light blue. In the case of PKS~1502$+$036, the detection with \emph{Swift} \cite{ABDO3} is indicated with an open blue circle.}
\label{fig:xrays}
\end{center}
\end{figure}

\subsection{The case of PMN~J0948$+$0022}
The case of PMN~J0948$+$0022 is worth underlining: there are several observations performed at the time of the discovery of the $\gamma$-ray emission and during the 2009 and 2010 MW Campaigns \cite{ABDO1,ABDO2,ABDO3,FOS3}. This offers the possibility to look at the source variations on months time scales and is a reference to understand how much a $\gamma$-NLS1 could change its position along the graphics prepared with MW data described above. To prepare the data, I considered as reference the time of the \emph{Swift} observations, from which I can measure X-ray and optical fluxes. \emph{Fermi} information were extracted from a 4-days long subset of LAT data, centered on the \emph{Swift} observation. The radio fluxes were extrapolated from the closest multifrequency observations by assuming a flat spectrum ($\alpha_{\rm r}=0$). 

A caveat should be noted when comparing optical-to-$\gamma$-ray data with radio measurements: indeed, the production of $\gamma$ rays require a compact region, which in turn is also optically thick at radio frequencies \cite{BLANDFORD2}. When the blob expands as it moves outward, then it becomes optically thin at radio frequencies, but its $\gamma$-ray production drops. The time lag between the flux peaks at $\gamma$ rays and radio frequencies is of the order of two months for PMN~J0948$+$0022 \cite{ABDO2,FOS3}, comparable with the lag observed in blazars (e.g. \cite{KOVALEV}). For the sake of simplicity, I used the radio measurements closest to the \emph{Swift} observations to prepare the graphics presented here, but -- as I explained -- it is necessary to take into account their different origin. This caveat is relaxed when dealing with the complete sample of NLS1s, because the observations at different wavelength have been done during specific surveys, which were not linked in time.

\section{Analysis of the sample}
With all the available data described in the previous Section, I prepared the graphics displayed in Fig.~\ref{fig:sample1} and Figs.~\ref{fig:radio}, \ref{fig:optical}, \ref{fig:xrays}. I searched for correlations of $\nu L_{\nu}$ at 100~MeV with its corresponding values at 1.4~GHz, 440~nm, and 1~keV, by using the \texttt{ASURV Rev.~1.2} software package \cite{ASURV1}, which makes use of the algorithms described in detail in \cite{FEIGELSON,ISOBE}. Nothing significant was found. When using only the $\gamma$-ray detections, there is something interesting, particularly at optical frequencies (Spearman's $\rho=0.82$, $P=0.044$), but 7 points are not sufficient for a robust correlation. It is necessary to repeat the analysis when more detections will be available. When adding the 34 bright blazars, it is possible to find significant correlations, but it is obviously due to the weight of the latter sample, where correlations are already known (e.g. \cite{ARSHAKIAN}).

No clear trend is visible in the FWHM H$\beta$ of the $\gamma$-NLS1s with respect to the other NLS1s of the sample, as well as in the radio loudness, except for the fact that all the $\gamma$-NLS1s are radio-loud (Fig.~\ref{fig:sample1}). The least radio-loud $\gamma$-NLS1 is FBQS~J1102$+$2239, with $R=32$. The source is rather unknown, but we could expect some degree of variability in the radio loudness, as shown by the changes of the points of PMN~J0948$+$0022 (maximum variation by a factor 2.6).

\begin{table}[!t]
\scriptsize
\begin{center}
\begin{tabular}{lcccccccc}
\hline
Quantities & \multicolumn{4}{c}{NLS1s} & \multicolumn{4}{c}{Blazars $+$ NLS1s}\\
\hline
{}         & \multicolumn{2}{c}{Kendall} & \multicolumn{2}{c}{Spearman} & \multicolumn{2}{c}{Kendall} & \multicolumn{2}{c}{Spearman}\\
\hline
{}         & $Z$ & $P$ & $\rho$ & $P$ & $Z$ & $P$ & $\rho$ & $P$\\
\hline
$L_{100~\rm MeV}$ vs $L_{1.4~\rm GHz}$ & 3.5 & 5$\times 10^{-4}$ & 0.59 & $< 10^{-4}$ & 9.2 & $< 10^{-4}$ & 0.86 & $< 10^{-4}$ \\ 
$L_{100~\rm MeV}$ vs $L_{440~\rm nm}$ & 1.2 & 0.23 & 0.30 & $9\times 10^{-3}$ & 8.2 & $< 10^{-4}$ & 0.75 & $< 10^{-4}$ \\ 
$L_{100~\rm MeV}$ vs $L_{1~\rm keV}$ & 1.9 & 0.052 & 0.27 & 0.02 & 9.1 & $< 10^{-4}$ & 0.76 & $< 10^{-4}$ \\ 
$L_{100~\rm MeV}$ vs $RL$  & 3.4 & 8$\times 10^{-4}$ & 0.49 & $< 10^{-4}$ & 8.5 & $< 10^{-4}$ & 0.81 & $< 10^{-4}$ \\ 
\hline
\end{tabular}
\end{center}
\normalsize
\caption{Summary of the correlation analysis. The number of points for the NLS1s sample is 76, with 69 upper limits at $\gamma$ rays and 28 upper limits at X-rays. The number of points of the Blazars $+$ NLS1s sample is 110, with the same number of upper limits. $P$ is the chance probability.}
\label{tab:asurv}
\end{table}

Among the faintest radio sources, 1H~0323$+$342 is the closest $\gamma$-NLS1, so in this case the low radio luminosity is compensated by the proximity of the source. Indeed, its flux density is not negligible ($\sim 600$~mJy; see Table~\ref{table:sample}). The two other sources, FBQS~J1102$+$2239 and SDSS~J124634.65$+$023809.0, are farther, with $z=0.455$ and $0.362$, respectively. Their fluxes are weak: $\sim 2$ and $\sim 38$~mJy, respectively (see Table~\ref{table:sample}), so there is some doubt that the $\gamma$-ray detection could be due to chance coincidence(\footnote{SDSS~J124634.65$+$023809.0 is also the weakest detection at $\gamma$-rays with $TS=15$, i.e. $\sim 4\sigma$. See Table~\ref{tab:latdetections}.}). These sources are anyway radio-loud, clearly because they are much more faint at optical frequencies. It is worth noting a bright \emph{Fermi} blazar close to the position of SDSS~J124634.65$+$023809.0 in the $\gamma$-ray vs radio loudness graphic (Fig.~\ref{fig:sample1}, \emph{right panel}), which is OX~$+$169 ($z=0.21$), although in term of radio power the difference between the two is about one order of magnitude (see Fig.~\ref{fig:radio}). There are other blazars with such low radio fluxes, at mJy level \cite{MASSARO}, some of them detected at high-energy $\gamma$ rays \cite{ABDO4}. Therefore, although the radio fluxes of the two NLS1s are low, it is not unlikely that they can generate high-energy photons, as it happens for blazars with comparable fluxes. 

I tried to estimate the brightness temperature, by using both the approaches based on the spectral characteristics \cite{READHEAD} and the variability at radio frequencies \cite{WAGNER}. In the latter case, there are two measurements done by the FIRST \cite{BECKER} and by the \emph{NRAO VLA Sky Survey} (NVSS, \cite{CONDON}), which were done with a few years of time separation. In the case of FBQS~J1102$+$2239, there is a change of about 67\% between the measurement performed during the NVSS (1993 Dec 6, 2.7$\pm$0.5~mJy) and FIRST (1996 Jan, 1.8$\pm$0.1~mJy), which results in an estimation of the brightness temperature of $\sim 10^{11}$~K. Instead, SDSS~J124634.65$+$023809.0 does not show evident indications of variability, although the two measurements were done on much larger timescale (NVSS, 36$\pm$1~mJy, 1995 Feb 27; FIRST, 38$\pm$1~mJy, 1998 July). The brightness temperature is then $\sim 5\times 10^{10}$~K. The estimation following \cite{READHEAD} and assuming a rather flat spectral index of $\alpha=0.1$, gives a temperature of $\sim 7\times 10^{10}$~K in both cases, which is comparable with the values obtained with the variability. This value obviously increases to exceed $10^{11}$~K if the spectral index becomes inverted. These temperatures are not outstandingly indicative of beamed non-thermal emission, but there is anyway some room for hope. Further observations are needed to better understand these sources and L. Fuhrmann, E. Angelakis et al. are planning multifrequency pointings with the Effelsberg radio telescope.

Figs.~\ref{fig:optical} and \ref{fig:xrays} display the $\gamma$ rays vs optical and X-ray data. As previously stated, there is some hint of correlation between optical and $\gamma$ rays when only the $\gamma$-NLS1s are considered. This can be explained having in mind the behavior observed in the 2009 MW Campaign of PMN~J0948$+$0022 \cite{ABDO2}, where a change in the optical emission due to the synchrotron emission was observed. This is confirmed by the recent polarization measurements in the $V$ filter performed by the Kanata telescope, which found a 19\% of polarization, comparable with the greatest value of the brightest blazars \cite{KANATA}. Also the optical emission of 1H~0323$+$342 is polarized, but with a lower degree (0.7$-$0.8\%), although still in the range of other blazars \cite{KANATA}(\footnote{In \cite{KANATA}, the two $\gamma$-NLS1s are erroneously classified as FSRQs.}). 

The fact that the $\gamma$-NLS1s are not clustered around some extreme values of $L_{1.4~\rm GHz}$, $L_{440~\rm nm}$, $L_{1~\rm keV}$, and radio loudness makes it clear that the measurements of high-energy $\gamma$ rays is still strongly dependent on the activity of the source. Indeed, PMN~J0948$+$0022 is the most active $\gamma$-NLS1 of these years and SBS~0846$+$513 was detected only recently, because of an increase of activity. This means that more discoveries are expected from a continuous monitoring of NLS1s.

\section{Host galaxy}
It is well-known the paradigm that associates the presence of powerful relativistic jets with elliptical host galaxies, while radio-quiet AGNs are hosted by spirals or even ellipticals when at high redshift (e.g. \cite{BAHCALL,KIRHAKOS,LAOR,SIKORA}). Some researchers suggested that this could be due to a requirement of great masses in the jet formation, where AGNs with jets need of $M>10^{9}M_{\odot}$ and those without jets have $M<3\times 10^{8}M_{\odot}$ (\cite{LAOR}, but see also \cite{FRANCESCHINI}). The former are common in ellipticals, while the latter are present in spirals.

The jet-elliptical paradigm was also the basis for elaborating a blazar evolution sequence, where highly-accreting and distant quasars evolve into lowly-accreting nearby BL Lac objects as the fueling of the central supermassive black hole is going to finish \cite{BOTTCHER,CAVALIERE,MARASCHI}. 

Although Ho \& Peng \cite{HO} have demonstrated that many Seyferts (hosted by spirals) become radio-loud once the nuclear luminosity is correctly measured, the counterexamples to this paradigm were extremely rare (e.g. \cite{BRUNTHALER1,BRUNTHALER2}; see the next Section). Today, it is clear that almost no Seyfert is really radio-silent, but there is more or less frequently some radio emission that can be interpreted as the basis of a jet or an outflow \cite{HO2,GIROLETTI2}. However, there is no general consensus on a specific explanation. Most important, none of these source was never detected at high-energy $\gamma$ rays: the recent detection with \emph{Fermi} of GeV $\gamma$ rays from the Seyfert 2s NGC~1068 and NGC~4945 can be explained with the emission from the starburst component and the proximity of the two sources \cite{LENAIN}(\footnote{Although, NGC~1068 might have also some contribution from a jet \cite{LENAIN}, but more observations are needed to better understand this source.}).

\begin{figure}[!t]
\begin{center}
\includegraphics[scale=0.5]{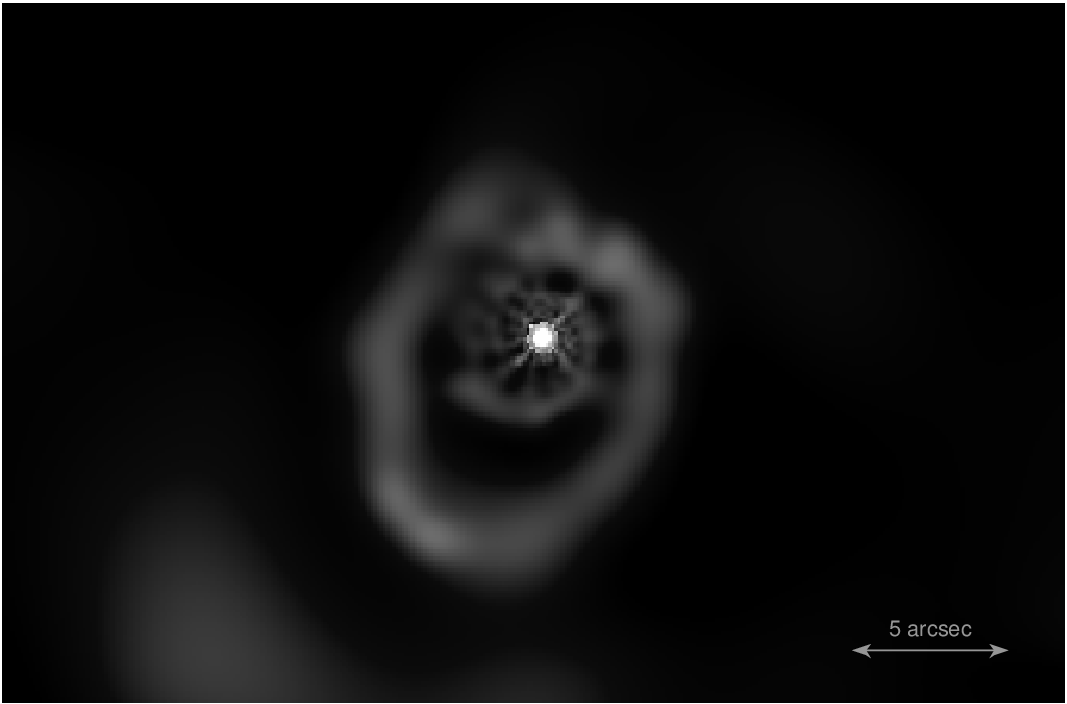}
\caption{\emph{Hubble Space Telescope}/WFPC2 image of 1H~0323$+$342 ($z=0.061$) with the filter F702W ($690$~nm). Two observations 200~s long were integrated and then adaptively smoothed \cite{EBELING} to emphasize the low brightness structures.}
\label{fig:hst}
\end{center}
\end{figure}

Some more cases of interesting flat spectrum and high brightness temperature radio emission were found in NLS1s (generally hosted by spirals), as reminded in the Introduction, and indeed these hints of powerful relativistic jets were then confirmed by the detection of GeV photons from a handful of NLS1s. Among the $\gamma$-NLS1s, only one can be imaged with sufficient resolution to understand its host galaxy: it is 1H~0323$+$342 ($z=0.061$), which was observed by the \emph{Hubble Space Telescope}. These observations have shown a spiral arm \cite{ZHOU3}, although other observations from the Nordic Optical Telescope (NOT) seem to suggest a circum-nuclear starburst ring, probably the trail of a recent merger \cite{ANTON}. The two options are not in contrast, because these starburst rings are common among NLS1s \cite{DEO} and the star forming activity is stronger in NLS1s than in other Seyferts \cite{SANI}. I have downloaded the \emph{HST} data and performed some image analysis, which resulted in an emphasis of the low brightness structures. As shown in Fig.~\ref{fig:hst}, the spiral structure is now more evident. 

Although no high-resolution observations are available for the remaining $\gamma$-NLS1s, it is likely that they are hosted by spirals, because this type of galaxy hosts almost all the low redshift NLS1s \cite{CRENSHAW,DEO}. Moreover, the presence of pseudobulges -- which are slowly built from spiral discs -- is in favor of the hypothesis that NLS1s are young active nuclei still growing by secular evolution(\footnote{Indeed, no NLS1 has been found a $z\gtrsim 0.8$.}) \cite{MATHUR2,ORBAN}. The intense starburst activity can be the result of a high accretion rate, supported by the large presence of bars, which in turn produce shocks in the interstellar medium and hence trigger the star formation. The presence of powerful relativistic jets might be a consequence of this young environment, because the jet can enhance the accretion, by draining angular momentum \cite{JOLLEY}. So, $\gamma$-NLS1s might be the low-redshift analog of high-redshift young quasars. The latter evolve into BL Lac Objects at low-$z$. What will be the fate of $\gamma$-NLS1s? Is it still to be written?

\begin{table}[!t]
\scriptsize
\begin{center}
\begin{tabular}{lrrccccccc}
\hline
Name & RA & Dec & z & Host & Ref. & $F_{0.1-100~\rm GeV}$ & $\Gamma$ & TS & err(dist)\\
\hline
III~Zw~2 & 2.6283 & $+$10.9703 & 0.090 & Sab? & \cite{BRUNTHALER1,BRUNTHALER2} & $<$2.0 & {} & {} & {}\\ 
NGC~612 & 23.4906 & $-$36.4932 & 0.030 & Sa & \cite{EMONTS2} & $<0.14$ & {} & {} & {}\\
PMN~J0315$-$1906 & 48.9671 & $-$19.1123 & 0.067 & S0ab? & \cite{LEDLOW1,LEDLOW2} & $<$0.08 & {} & {} & {}\\
PKS~0336$-$177 & 54.8070 & $-$17.6002 & 0.065 & S0 & \cite{LOVEDAY} & 0.25$\pm$0.14 & 1.7$\pm$0.2 & 80 & 0.045(0.026)\\
3C~120 & 68.2962 & $+$5.3543 & 0.033 & S0 & \cite{MOLES} & $<$2.2 & {} & {} & {}\\
B2~0722$+$30 & 111.4057 & $+$29.9541 & 0.019 & S0 & \cite{EMONTS} & $<$0.12 & {} & {} & {}\\ 
3C~277.1 & 193.1098 & $+$56.5721 & 0.320 & S0? & \cite{HAMILTON1,HAMILTON2} & $<$0.03 & {} & {} & {}\\
PG~1309$+$355 & 198.0740 & $+$35.2559 & 0.183 & Sab & \cite{HAMILTON1,HAMILTON2} & $<2.3$ & {} & {} & {}\\ 
PKS~1413$+$135 & 213.9951 & $+$13.3399 & 0.247 & Sa & \cite{PERLMAN} & 4.1$\pm$0.6 & 3.0$\pm$0.1 & 119 & 0.075(0.051)\\
\hline
\end{tabular}
\end{center}
\normalsize
\caption{Sample of radio galaxies in spiral hosts. The coordinates are at the epoch J2000. The $\gamma$-ray flux is in units of [$10^{-8}$~ph~cm$^{-2}$~s$^{-1}$]; upper limits are at $5\sigma$ confidence level. The last column, indicating the 95\% error circle of the LAT source and the distance from the radio position, is in [deg]. PKS~0336$-$177 is also present in the first LAT catalog \cite{ABDO4}. 3C~120 is not present in the first LAT catalog \cite{ABDO4}, although a detection is reported in \cite{ABDO5}.}
\label{tab:radiogalaxies}
\end{table}

\section{Search for the parent population}
The presence of a population of AGNs with beamed jets poses the question of the parent population, i.e. those sources of the same type but viewed at random orientations. The number of the parent population is $\sim 2\Gamma^2$ times the number of beamed sources.  Therefore, if we know 7 $\gamma$-NLS1s, this means that there should be $\sim 1400$ unbeamed parent sources, having assumed the common value of $\Gamma=10$. The parent population of blazars is that of the radio galaxies, but what is the parent population of $\gamma$-NLS1s? 

The only known NLS1 displaying extended radio structures on kpc scale is PKS~0558$-$504, with a $\sim 46$~kpc bipolar jet detected at 4.8~GHz \cite{GLIOZZI}. The extreme compactness and lack of extended structures of the $\gamma$-NLS1s at radio frequencies \cite{DOI,YUAN,FOS3,GIROLETTI} might suggest that when observed at large angles, these sources simply becomes radio-quiet, because there is negligible unbeamed radio emission. In this case, the parent population of $\gamma$-NLS1s might simply be that of radio-quiet NLS1s. 

There could be another possibility. If the broad-line region (BLR) of NLS1s is disc-like and we are observing it pole-on, then there is no motion along the viewing angle, the Doppler broadening is negligible, and the profiles of the emission lines of the BLR are narrower than usual \cite{DECARLI}. This has been observed also when analyzing large samples of radio-loud AGNs and explained with a disc-like BLR viewed at small angles \cite{MCLURE}. In the case of NLS1s, this means that when observed at large angles, they become the usual broad-line Seyferts. Therefore, the parent population should be searched among the broad-line radio galaxies hosted by spirals, which are not very common \cite{VERON}. Only recently, Inskip et al. performed a near-IR morphological study on the 2~Jy sample of radio-loud AGNs and found that $\sim 12$\% of the sources are hosted by disc galaxies \cite{INSKIP}. In the past, the examples of radio-loud AGNs in spirals were just a few (e.g. \cite{HAMILTON1,HAMILTON2}). Others can be found in the literature, but they were a handful (see Table~\ref{tab:radiogalaxies}). It is worth noting that often these sources are hosted by S0 galaxies and, therefore, there could be the risk of misclassification: a faint elliptical galaxy can be erroneously classified as a bright S0, and vice versa. 

Therefore, the parent population of the $\gamma$-NLS1s is still basically unknown.

\section{Implications on the classification of AGNs}
Before the discovery of $\gamma$-NLS1s, the AGNs with relativistic jets viewed at small angles were called blazars. Blazars were divided into FSRQ and BL Lac Objects, depending on the equivalent width of their emission lines ($\gtrless 5$~\AA; e.g. \cite{URRY}), which in turn translates in an indication of the accretion power(\footnote{Recently, Ghisellini et al. \cite{GHISELLINI3} proposed to revise this classification on the basis of the line luminosities in Eddington units. In this case, the threshold is at $L_{\rm BLR}/L_{\rm Edd}\sim 5\times 10^{-4}$.}). The same type of host suggested an evolutionary link between the two types of sources. Therefore, the term blazar indicates an AGN with a relativistic jet viewed at small angles, hosted in an elliptical galaxy, with an accretion disc and a BLR more or less developed, depending on the luminosity of the disc. The quasars have high accretion ($\sim 0.1L_{\rm Edd}$) and fully developed BLR, while BL Lacs have low accretion ($<10^{-3}L_{\rm Edd}$) and weak BLR (e.g. \cite{GHISELLINI2}). The FWHMs of the lines were always $>2000$~km/s \cite{FINE,WILLS}. The masses of the central black hole are in the range of $10^{8-10}M_{\odot}$ (e.g. \cite{GHISELLINI2}). They were born as FSRQ and evolve into BL Lac Objects, as the fuel is ending \cite{BOTTCHER,CAVALIERE,MARASCHI}. Therefore, it is the same type of source (blazar) and the terms FSRQ and BL Lac Object refer to specific evolutionary stages, where the nearby environment can determine a more or less efficient cooling of the relativistic electrons of the jet \cite{GHISELLINI4}.

\begin{figure}[!t]
\begin{center}
\includegraphics[angle=270,scale=0.5]{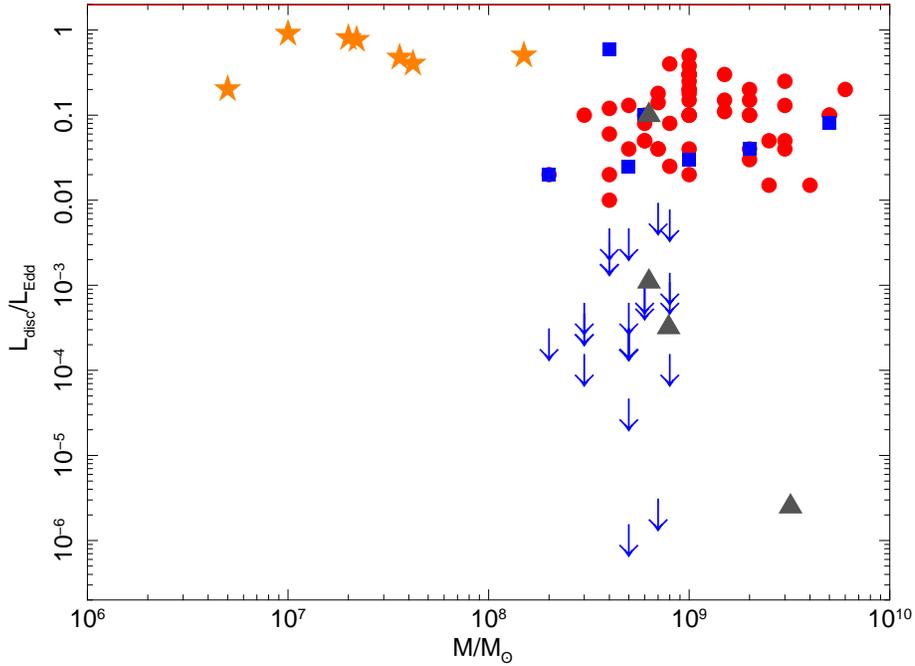}
\vskip 36pt
\caption{AGNs with relativistic jets in terms of mass and accretion power (in Eddington units). FSRQ are indicated with red circles and BL Lac Objects are represented with blue squares or arrows, in the case it was not possible to estimate the accretion power and there is only an upper limit (data from \cite{GHISELLINI2}). Some radio galaxies detected at $\gamma$ rays \cite{ABDO5} are indicated with gray triangles (data from \cite{SIKORA}). Orange stars are for $\gamma$-NLS1s (data for the first four are from \cite{ABDO3}, while the three newly detected are from the present work).}
\label{fig:massaccr}
\end{center}
\end{figure}

Some researchers use the term blazar to indicate only the emission from the relativistic jet viewed at small angles, because it overwhelms almost all the other components of the AGN. Sometimes, it is possible to read terms like ``quasar-hosted blazars'' (e.g. \cite{SIKORA2}). However, I think that this terminology is misleading for a couple of reasons: it is rather evident that several important informations are lost, specifically on the AGN components and hosts, and therefore it is necessary to add other words to recover this information, as in the above example. Moreover, there could be confusion with relativistic jets in Galactic compact objects: if the term blazar should indicate only the emission from a jet viewed at small angles, then also microquasars should be simply named blazars. 

Now, with the arrival of $\gamma$-NLS1s, the problems are increasing. Indeed, if we continue naming blazars any AGN with a relativistic jet viewed at small angles, we should call in this way also the $\gamma$-NLS1s, thus losing a lot of informations and questions. $\gamma$-NLS1s are different from blazars in a lot of key aspects: smaller masses ($10^{6-8}M_{\odot}$), greater accretion rates ($>>0.1L_{\rm Edd}$, up to the Eddington value), smaller FWHMs of the BLR lines ($<$~2000~km/s), different hosts (spiral). Surely the jet in $\gamma$-NLS1s is very similar to the jets in blazars (see Fig.~1 in \cite{FOS2}), but everything else is different. Indeed, as we plot a graphic with accretion power vs central mass, then the difference between $\gamma$-NLS1s, blazars and radio galaxies is immediately evident (see Fig.~\ref{fig:massaccr}).

Therefore, I think it is better to keep the name blazars for FSRQ plus BL Lacs, so that this term refers to a global set of properties of the whole AGN and its host, rather than only of the jet. To summarize, we could then divide the AGNs with relativistic jets according to the Table~\ref{tab:classification}.

\begin{table}[!t]
\begin{center}
\begin{tabular}{ll|rr}
\hline
\multicolumn{2}{c}{\emph{Beamed}} & \multicolumn{2}{c}{\emph{Unbeamed (parent population)}}\\
\hline
Blazars: & {}      & Radio Galaxies: & {}\\
{}      & - FSRQ    & {} & - High-Excitation RG\\
{}      & - BL Lacs & {} & - Low-Excitation RG\\
\hline
\multicolumn{2}{l}{$\gamma$-NLS1s} & \multicolumn{2}{l}{RGs in spirals or misaligned NLS1s?} \\
\hline
\end{tabular}
\end{center}
\caption{Tentative classification of AGNs with relativistic jets. For the definition of High- and Low-Excitation RG see, for example, \cite{BUTTIGLIONE}.}
\label{tab:classification}
\end{table}

The evidence of powerful relativistic jets in NLS1s confirmed that these structures are rather ubiquitous in the Universe and that the jet-elliptical paradigm was an observational bias. It is worth reminding the exchange of words by R. Blandford and G. Burbidge in the discussion after Blandford's seminal talk at the ``Pittsburgh Conference on BL Lac Objects'' in 1978 \cite{BLANDFORD}:

\begin{quote}
G. Burbidge: ``Roger, is your model for AO~0235$+$164 an elliptical galaxy with a black hole at the center?''\\
R. Blandford: ``I would say that it is either a galaxy or a proto-galaxy. As the continuum emission is proposed to originate in the central 10 pc, \emph{I don't think the nature of the surrounding object is particularly relevant to the model}.'' [the emphasis is mine].
\end{quote}

\section{Prospects}
I prefer to write something on the future, rather than to write a conclusion section. Indeed, the discovery of $\gamma$-NLS1s opens many more questions rather than to solve any single doubt. Something has been done, but much more has still to be done. I have no answers, but a lot of questions. In my opinion, the key points in this research field are:

\begin{itemize}
\item The first point is rather obvious: improve the sample of $\gamma$-NLS1s! As the detection depends on the activity of the source, the monitoring of the sky with \emph{Fermi} guarantees an open space for new discoveries. With more sources, it will be possible to search for correlations and better assess the statistical properties of $\gamma$-NLS1s.

\item Study the time-resolved MW properties of $\gamma$-NLS1s: a few detailed cases study can often offer many more and deeper informations on the nature of certain cosmic sources than a statistical study on a wide sample. 

\item Study the host galaxy: to date, only one $\gamma$-NLS1 is known to be hosted in a spiral galaxy (1H~0323$+$342), while for the others we can infer the same host type. Obviously, one thing is the inference, another thing is the direct observation.

\item Search for the parent population: in this case, it is not yet clear if we have to search for other NLS1s with kpc structures, like PKS~0558$-$504, or if the answer is among the broad-line radio galaxies in spiral hosts. 
\end{itemize}

\newpage
\appendix

\section{List of NLS1}
The sample of 76 NLS1s is reported in the following Table~\ref{table:sample}. The MW data are those observed (Table~\ref{table:obs}), i.e. not corrected, except for $R$, which has been calculated after having dereddened the optical flux.

\begin{table}[!h]
\scriptsize
\renewcommand\thetable{A1}
\begin{center}
\begin{tabular}{lrrccrr}
\hline
Name & RA & Dec & $z$ & $N_{\rm H}$ & FWHM~H$\beta$ & $R$\\
{} & J2000 & J2000 & {} & [$10^{20}$~cm$^{-2}$] & [km/s] & {}\\
\hline
FBQS~J0022$-$1039 & 5.7051 & $-$10.6656 & 0.414 & 3.11 & 1845 & 6\\
FBQS~J0100$-$0200 & 15.1342 & $-$2.0128 & 0.227 & 4.05 & 920 & 8 \\
PMN~J0134$-$4258  & 23.5704 & $-$42.9742 & 0.238 & 1.69 & 930 & 209 \\
1H~0323$+$342 & 51.1715 & $+$34.1794 & 0.061 & 11.7 & 1600 & 318 \\
PKS~0558$-$504 & 89.9474 & $-$50.4478 & 0.137 & 3.46 & 1500 & 27 \\
FBQS~J0706$+$3901 & 106.6048 & $+$39.0310 & 0.086 & 8.07 & 664 & 7 \\
FBQS~J0713$+$3820 & 108.4179 & $+$38.3444 & 0.123 & 6.00 & 1487 & 20 \\
FBQS~J0729$+$3046 & 112.4679 & $+$30.7792 & 0.150 & 6.00 & 891 & 2 \\
FBQS~J0736$+$3926 & 114.0964 & $+$39.4383 & 0.118 & 5.76 & 1806 & 3 \\
FBQS~J0744$+$5149 & 116.0096 & $+$51.8216 & 0.460 & 4.83 & 1989 & 59 \\
FBQS~J0752$+$2617 & 118.1900 & $+$26.2933 & 0.082 & 3.42 & 1906 & 2 \\
FBQS~J0758$+$3920 & 119.5002 & $+$39.3414 & 0.095 & 5.12 & 1908 & 90 \\
FBQS~J0804$+$3853 & 121.0385 & $+$38.8969 & 0.151 & 5.16 & 1356 & 10 \\
RGB~J0806$+$728   & 121.6624 & $+$72.8057 & 0.098 & 2.98 & 1060 & 41 \\
FBQS~J0810$+$2341 & 122.6086 & $+$23.6989 & 0.133 & 4.50 & 1831 & 6 \\
RGB~J0814$+$561   & 123.6338 & $+$56.1657 & 0.509 & 4.49 & 2164 & 361 \\
FBQS~J0818$+$3834 & 124.7053 & $+$38.5711 & 0.160 & 3.84 & 1683 & 6 \\
SBS~0846$+$513    & 132.4916 & $+$51.1414 & 0.584 & 3.00 & 1811 & 4496 \\
SDSS~J085001.17$+$462600.5 & 132.5049 & $+$46.4335 & 0.523 & 2.64 & 1251 & 318\\
PMN~J0902$+$0442 & 135.6132 & $+$4.7193 & 0.532 & 3.11 & 2089 & 1974 \\
FBQS~J0909$+$3124 & 137.4494 & $+$31.4121 & 0.265 & 1.74 & 1610 & 8 \\
FBQS~J0937$+$3615 & 144.2626 & $+$36.2603 & 0.180 & 1.21 & 1048 & 12 \\
FBQS~J0946$+$3223 & 146.5458 & $+$32.3906 & 0.405 & 1.52 & 1615 & 5 \\
PMN~J0948$+$0022 & 147.2388 & $+$0.3738 & 0.585 & 5.20 & 1432 & 846 \\
SDSS~J095317.09$+$283601.5 & 148.3212 & $+$28.6004 & 0.657 & 1.28 & 2162 & 665 \\
RBS~826 & 151.4244 & $+$43.5445 & 0.179 & 1.04 & 2059 & 4  \\
FBQS~J1010$+$3003 & 152.5029 & $+$30.0560 & 0.256 & 2.35 & 1305 & 2 \\
SDSS~J103123.73$+$423439.3 & 157.8489 & $+$42.5776 & 0.376 & 0.99 & 1642 & 291 \\
SDSS~J103727.45$+$003635.6 & 159.3644 & $+$0.6099 & 0.595 & 4.87 & 1357 & 569 \\
FBQS~J1038$+$4227 & 159.7483 & $+$42.4617 & 0.220 & 1.49 & 1979 & 10 \\
B3~1044$+$476 & 161.8861 & $+$47.4256 & 0.798 & 1.30 & 2153 & 12281 \\
RX~J1048.3$+$2222 & 162.0691 & $+$22.3775 & 0.329 & 1.46 & 1301 & 10 \\
FBQS~J1102$+$2239 & 165.5974 & $+$22.6557 & 0.455 & 1.21 & 1972 & 32 \\
1107$+$372 & 167.5210 & $+$36.8934 & 0.630 & 1.69 & 1300 & 1393 \\
B2~1111$+$32 & 168.6621 & $+$32.6926 & 0.189 & 1.86 & 1980 & 1986\\
FBQS~J1127$+$2654 & 171.9016 & $+$26.9140 & 0.379 & 1.36 & 1903 & 4\\
FBQS~J1136$+$3432 & 174.2331 & $+$34.5436 & 0.193 & 1.71 & 918 & 4\\
SDSS~J113824.54$+$365327.1 & 174.6023 & $+$36.8908 & 0.356 & 1.75 & 1364 & 328\\
SDSS~J114654.28$+$323652.3 & 176.7262 & $+$32.6145 & 0.465 & 1.42 & 2081 & 154\\
FBQS~J1159$+$2838 & 179.8222 & $+$28.6374 & 0.209 & 1.65 & 1415 & 20\\
FBQS~J1220$+$3853 & 185.1463 & $+$38.8879 & 0.377 & 2.21 & 1917 & 4\\
FBQS~J1227$+$3214 & 186.9548 & $+$32.2497 & 0.137 & 1.62 & 951 & 91\\
SDSS~J123852.12$+$394227.8 & 189.7173 & $+$39.7077 & 0.622 & 1.46 & 910 & 268\\
SDSS~J124634.65$+$023809.0 & 191.6444 & $+$2.6358 & 0.362 & 2.01 & 1425 & 277\\
FBQS~J1256$+$3852 & 194.0086 & $+$38.8752 & 0.419 & 1.73 & 2079 & 8\\
SDSS~J130522.75$+$511640.3 & 196.3448 & $+$51.2778 & 0.785 & 0.98 & 1925 & 277\\
FBQS~J1333$+$4141 & 203.4395 & $+$41.6910 & 0.225 & 0.74 & 1940 & 9\\
FBQS~J1346$+$3121 & 206.6457 & $+$31.3594 & 0.246 & 1.27 & 1600 & 11\\
\hline
\end{tabular}
\end{center}
\caption{Sample of NLS1s.}
\label{table:sample}
\normalsize
\end{table}

\begin{table}[!t]
\scriptsize
\renewcommand\thetable{A1}
\begin{center}
\begin{tabular}{lrrccrr}
\hline
Name & RA & Dec & $z$ & $N_{\rm H}$ & FWHM~H$\beta$ & $R$ \\
\hline
FBQS~J1358$+$2658 & 209.6891 & $+$26.9690 & 0.331 & 1.55 & 1863 & 11\\
FBQS~J1405$+$2555 & 211.3176 & $+$25.9261 & 0.165 & 1.29 & 1398 & 1\\
FBQS~J1408$+$2409 & 212.1159 & $+$24.1569 & 0.131 & 1.57 & 1590 & 4\\
FBQS~J1421$+$2824 & 215.3086 & $+$28.4145 & 0.540 & 1.28 & 1838 & 204 \\
SDSS~J143509.49$+$313147.8 & 218.7897 & $+$31.5301 & 0.501 & 1.14 & 1719 & 949 \\
FBQS~J1442$+$2623 & 220.6700 & $+$26.3924 & 0.108 & 2.13 & 795 & 5 \\
B3~1441$+$476 & 220.8273 & $+$47.4324 & 0.703 & 1.47 & 1848 & 1067 \\
FBQS~J1448$+$3559 & 222.1046 & $+$35.9963 & 0.114 & 1.00 & 1856 & 2 \\
PKS~1502$+$036 & 226.2770 & $+$3.4419 & 0.408 & 3.89 & 1082 & 3364 \\
FBQS~J1517$+$2239 & 229.3844 & $+$22.6564 & 0.109 & 3.73 & 1789 & 6 \\
FBQS~J1519$+$2838 & 229.9006 & $+$28.6410 & 0.270 & 2.28 & 1641 & 4 \\
RGB~J1548$+$351 & 237.0747 & $+$35.1912 & 0.478 & 2.28 & 2035 & 701 \\
FBQS~J1612$+$4219 & 243.2493 & $+$42.3279 & 0.233 & 1.19 & 819 & 24 \\
RGB~J1629$+$401 & 247.2555 & $+$40.1332 & 0.272 & 1.05 & 1260 & 50 \\
RGB~J1633$+$473 & 248.3483 & $+$47.3164 & 0.116 & 1.79 & 909 & 154 \\
SDSS~J163401.94$+$480940.2 & 248.5081 & $+$48.1612 & 0.494 & 1.64 & 1609 & 187 \\
RGB~J1644$+$263 & 251.1772 & $+$26.3203 & 0.144 & 5.12 & 1507 & 396 \\
FBQS~J1702$+$3247 & 255.6294 & $+$32.7888 & 0.164 & 1.71 & 1400 & 1 \\
B3~1702$+$457 & 255.8766 & $+$45.6798 & 0.060 & 2.52 & 490 & 102 \\
FBQS~J1713$+$3523 & 258.2686 & $+$35.3926 & 0.085 & 2.45 & 1002 & 10 \\
FBQS~J1716$+$3112 & 259.0081 & $+$31.2038 & 0.110 & 3.00 & 1571 & 1 \\
FBQS~J1718$+$3042 & 259.7096 & $+$30.7004 & 0.281 & 3.06 & 1434 & 3 \\
SDSS~J172206.03$+$565451.6 & 260.5251 & $+$56.9143 & 0.425 & 2.10 & 1385 & 323 \\
PKS~2004$-$447 & 301.9799 & $-$44.5790 & 0.240 & 2.96 & 1447 & 6358 \\
PHL~1811 & 328.7563 & $-$9.3734 & 0.192 & 4.04 & 1500 & 1 \\
RX~J2159.4$+$0113 & 329.8502 & $+$1.2182 & 0.101 & 4.27 & 1429 & 3 \\
FBQS~J2327$-$1023 & 351.8105 & $-$10.3882 & 0.065 & 2.25 & 652 & 1 \\
FBQS~J2338$-$0900 & 354.6415 & $-$9.0109 & 0.374 & 2.21 & 1564 & 9 \\
\hline
\end{tabular}
\end{center}
\caption{-- \emph{Continue}.}
\normalsize
\end{table}

\newpage

\begin{table}[!t]
\scriptsize
\renewcommand\thetable{A2}
\begin{center}
\begin{tabular}{lcccc}
\hline
Name & $f_{1.4~\rm GHz}$ & $m_{B}$ & $F_{0.1-2.4~\rm keV}$ & $F_{0.1-100~\rm GeV}$\\
{}   & [mJy] & [mag] & [$10^{-12}$~erg~cm$^{-2}$~s$^{-1}$] & [$10^{-8}$~ph~cm$^{-2}$~s$^{-1}$]\\
\hline
FBQS~J0022$-$1039 & 1.9$\pm$0.1 & 17.81$\pm$0.05 & $<$0.2 & $<$0.15\\
FBQS~J0100$-$0200 & 5.9$\pm$0.1 & 17.1$\pm$0.1 & $<$0.2 & $<$0.11\\
PMN~J0134$-$4258  & 55$\pm$9 & 18.0$\pm$0.3 & 2.4$\pm$0.3 & $<$2.6\\
1H~0323$+$342 &  613$\pm$21 & 16.38$\pm$0.03 & 8.0$\pm$0.7 & 6.0$\pm$0.7\\
PKS~0558$-$504 &  121$\pm$10 & 15.08$\pm$0.01 & 88$\pm$6 & $<$2.4\\
FBQS~J0706$+$3901 &  4.3$\pm$0.5 & 17.6$\pm$0.3 & $<$0.3 & $<$0.15\\
FBQS~J0713$+$3820 &  10.8$\pm$0.1 & 15.1$\pm$0.3 & 3.2$\pm$0.5 & $<$3.1\\
FBQS~J0729$+$3046 &  1.2$\pm$0.1 & 17.4$\pm$0.3 & 0.50$\pm$0.22 & $<$2.1\\
FBQS~J0736$+$3926 &  3.6$\pm$0.1 & 16.53$\pm$0.06 & 9.0$\pm$0.8 & $<$0.16\\
FBQS~J0744$+$5149 &  11.9$\pm$0.1 & 18.4$\pm$0.3 & 1.2$\pm$0.3 & $<$0.084\\
FBQS~J0752$+$2617 &  1.3$\pm$0.1 & 17.06$\pm$0.05 & 3.3$\pm$0.4 & $<$2.0\\
FBQS~J0758$+$3920 &  11.6$\pm$0.1 & 19.08$\pm$0.08 & 2.0$\pm$0.5 & $<$3.1\\
FBQS~J0804$+$3853 &  2.7$\pm$0.1 & 18.27$\pm$0.08 & $<$0.3 & $<$0.066\\
RGB~J0806$+$728   &  50.1$\pm$0.2 & 16.5$\pm$0.3 & 2.9$\pm$0.6 & $<$1.7\\
FBQS~J0810$+$2341 &  0.9$\pm$0.1 & 18.89$\pm$0.07 & 0.99$\pm$0.35 & $<$4.2\\
RGB~J0814$+$561   &  80.2$\pm$0.2 & 18.27$\pm$0.04 & 1.2$\pm$0.2 & $<$1.4\\
FBQS~J0818$+$3834 &  2.1$\pm$0.1 & 17.92$\pm$0.06 & $<$0.2 & $<$3.7\\
SBS~0846$+$513    &  350.0$\pm$0.1 & 19.27$\pm$0.07 & 0.23$\pm$0.03 & 0.51$\pm$0.15\\
SDSS~J085001.17$+$462600.5 &  21.3$\pm$0.1 & 19.43$\pm$0.06 & $<$0.18 & $<$3.4\\
PMN~J0902$+$0442 &  156.5$\pm$0.2 & 19.28$\pm$0.06 & $<$0.2 & $<$2.7\\
FBQS~J0909$+$3124 &  1.5$\pm$0.1 & 18.35$\pm$0.08 & $<$0.14 & $<$1.3\\
FBQS~J0937$+$3615 &  3.2$\pm$0.1 & 17.99$\pm$0.07 & 1.2$\pm$0.2 & $<$2.1\\
FBQS~J0946$+$3223 &  1.6$\pm$0.1 & 17.81$\pm$0.05 & $<$0.13 & $<$7.0\\
PMN~J0948$+$0022 &  111.5$\pm$0.1 & 18.86$\pm$0.05 & 1.0$\pm$0.3 & 13.7$\pm$0.7\\
SDSS~J095317.09$+$283601.5 & 47.9$\pm$0.2 & 19.21$\pm$0.04 & $<$0.12 & $<$1.4\\
RBS~826 &  2.8$\pm$0.1 & 16.87$\pm$0.05 & 7.4$\pm$0.5 & $<$0.39 \\
FBQS~J1010$+$3003 &  1.0$\pm$0.1 & 17.23$\pm$0.05 & 2.2$\pm$0.3 & $<$6.5\\
SDSS~J103123.73$+$423439.3 &  17.0$\pm$0.1 & 19.52$\pm$0.07 & $<$0.1 & $<$3.1\\
SDSS~J103727.45$+$003635.6 &  27.9$\pm$0.1 & 19.91$\pm$0.06 & $<$0.25 & $<$2.7\\
FBQS~J1038$+$4227 &  2.4$\pm$0.1 & 18.11$\pm$0.07 & $<$0.13 & $<$0.071\\
B3~1044$+$476 &  789$\pm$24 & 19.29$\pm$0.06 & 0.27$\pm$0.11 & $<$0.78\\
RX~J1048.3$+$2222 &  1.5$\pm$0.1 & 18.62$\pm$0.06 & 1.3$\pm$0.3 & $<$0.38\\
FBQS~J1102$+$2239 &  1.8$\pm$0.1 & 19.55$\pm$0.07 & 0.62$\pm$0.26 & 2.0$\pm$0.6\\
1107$+$372 &  23.3$\pm$0.8 & 20.83$\pm$0.08 & $<$0.14 & $<$0.73\\
B2~1111$+$32 &  107.8$\pm$0.1 & 19.7$\pm$0.1 & $<$0.15 & $<$0.62\\
FBQS~J1127$+$2654 &  1.9$\pm$0.1 & 17.24$\pm$0.06 & $<$0.13 & $<$3.1\\
FBQS~J1136$+$3432 & 1.3$\pm$0.1 & 17.88$\pm$0.06 & $<$0.14 & $<$0.11\\
SDSS~J113824.54$+$365327.1 &  12.6$\pm$0.1 & 20.04$\pm$0.08 & $<$0.15 & $<$4.9\\
SDSS~J114654.28$+$323652.3 &  15.4$\pm$0.1 & 18.93$\pm$0.05 & 0.22$\pm$0.10 & $<$5.1\\
FBQS~J1159$+$2838 &  2.0$\pm$0.1 & 19.04$\pm$0.07 & $<$0.14 & $<$2.1\\
FBQS~J1220$+$3853 &  2.2$\pm$0.1 & 17.14$\pm$0.05 & 0.41$\pm$0.16 & $<$1.5\\
FBQS~J1227$+$3214 &  6.4$\pm$0.1 & 19.5$\pm$0.1 & 0.98$\pm$0.28 & $<$0.41\\
SDSS~J123852.12$+$394227.8 &  11.2$\pm$0.1 & 19.84$\pm$0.05 & 0.66$\pm$0.13 & $<$0.98\\
SDSS~J124634.65$+$023809.0 &  38.1$\pm$0.1 & 18.67$\pm$0.05 & $<$0.16 & 1.7$\pm$0.7\\
FBQS~J1256$+$3852 &  2.1$\pm$0.1 & 17.90$\pm$0.05 & $<$0.14 & $<$0.15\\
SDSS~J130522.75$+$511640.3 &  86.9$\pm$0.1 & 17.54$\pm$0.05 & $<$0.11 & $<$1.6\\
FBQS~J1333$+$4141 &  2.0$\pm$0.1 & 18.10$\pm$0.06 & 0.49$\pm$0.10 & $<$2.9\\
FBQS~J1346$+$3121 &  1.4$\pm$0.1 & 18.77$\pm$0.06 & $<$0.12 & $<$0.087\\
FBQS~J1358$+$2658 &  1.2$\pm$0.1 & 18.95$\pm$0.07 & $<$0.14 & $<$2.0\\
\hline
\end{tabular}
\end{center}
\caption{Sample of NLS1s: \emph{observed} data (i.e. not corrected).}
\label{table:obs}
\normalsize
\end{table}

\begin{table}[!t]
\scriptsize
\renewcommand\thetable{A2}
\begin{center}
\begin{tabular}{lcccc}
\hline
Name & $f_{1.4~\rm GHz}$ & $m_{B}$ & $F_{0.1-2.4~\rm keV}$ & $F_{0.1-100~\rm GeV}$\\
{}   & [mJy] & [mag] & [$10^{-12}$~erg~cm$^{-2}$~s$^{-1}$] & [$10^{-8}$~ph~cm$^{-2}$~s$^{-1}$]\\
\hline
FBQS~J1405$+$2555 & 0.7$\pm$0.1 & 15.46$\pm$0.04 & 8.1$\pm$0.5 & $<$0.047\\
FBQS~J1408$+$2409 &  3.0$\pm$0.1 & 16.96$\pm$0.06 & 3.4$\pm$0.4 & $<$0.64\\
FBQS~J1421$+$2824 &  48.7$\pm$0.1 & 17.95$\pm$0.04 & $<$0.13 & $<$0.078\\
SDSS~J143509.49$+$313147.8 &  44.7$\pm$0.1 & 19.72$\pm$0.06 & 1.2$\pm$0.2 & $<$1.1\\
FBQS~J1442$+$2623 &  3.4$\pm$0.1 & 17.06$\pm$0.07 & 1.2$\pm$0.2 & $<$1.0\\
B3~1441$+$476 &  171.1$\pm$0.1 & 18.35$\pm$0.04 & 0.29$\pm$0.12 & $<$1.1\\
FBQS~J1448$+$3559 &  1.5$\pm$0.1 & 16.87$\pm$0.06 & 4.2$\pm$0.3 & $<$0.14\\
PKS~1502$+$036 &  380.5$\pm$0.1 & 18.99$\pm$0.06 & $<$0.22 & 7.0$\pm$0.6\\
FBQS~J1517$+$2239 &  1.1$\pm$0.1 & 18.56$\pm$0.07 & 0.43$\pm$0.19 & $<$1.0\\
FBQS~J1519$+$2838 &  2.0$\pm$0.1 & 17.34$\pm$0.06 & 0.65$\pm$0.21 & $<$1.9\\
RGB~J1548$+$351 &  141.5$\pm$0.1 & 18.22$\pm$0.05 & 0.92$\pm$0.31 & $<$1.4\\
FBQS~J1612$+$4219 &  3.6$\pm$0.1 & 18.56$\pm$0.06 & $<$0.12 & $<$0.14\\
RGB~J1629$+$401 &  11.9$\pm$0.2 & 18.04$\pm$0.05 & 8.8$\pm$0.3 & $<$1.2\\
RGB~J1633$+$473 &  65.0$\pm$0.1 & 17.55$\pm$0.06 & 2.8$\pm$0.3 & $<$0.097\\
SDSS~J163401.94$+$480940.2 &  7.7$\pm$0.1 & 19.89$\pm$0.06 & 0.20$\pm$0.08 & $<$1.0\\
RGB~J1644$+$263 &  90.8$\pm$0.2 & 18.40$\pm$0.06 & 2.8$\pm$0.5 & $<$1.9\\
FBQS~J1702$+$3247 &  1.5$\pm$0.1 & 16.15$\pm$0.05 & 8.1$\pm$0.4 & $<$2.8\\
B3~1702$+$457 &  118.6$\pm$0.1 & 16.52$\pm$0.08 & 16.1$\pm$0.7 & $<$1.3\\
FBQS~J1713$+$3523 &  11.2$\pm$0.1 & 16.56$\pm$0.06 & 17.2$\pm$0.7 & $<$0.45\\
FBQS~J1716$+$3112 &  2.4$\pm$0.1 & 16.18$\pm$0.05 & 5.4$\pm$0.4 & $<$1.8\\
FBQS~J1718$+$3042 &  0.6$\pm$0.1 & 18.45$\pm$0.06 & 0.70$\pm$0.17 & $<$0.68\\
SDSS~J172206.03$+$565451.6 &  39.8$\pm$0.1 & 18.76$\pm$0.06 & 1.3$\pm$0.2 & $<$1.3\\
PKS~2004$-$447 &  791$\pm$38 & 18.9$\pm$0.3 & 0.44$\pm$0.22 & 1.2$\pm$0.3\\
PHL~1811 &  1.2$\pm$0.1 & 13.9$\pm$0.3 & $<$0.22 & $<$2.1\\
RX~J2159.4$+$0113 &  2.1$\pm$0.2 & 17.28$\pm$0.06 & 1.4$\pm$0.5 & $<$5.7\\
FBQS~J2327$-$1023 &  2.2$\pm$0.1 & 16.02$\pm$0.07 & $<$0.17 & $<$2.2\\
FBQS~J2338$-$0900 &  1.8$\pm$0.1 & 18.29$\pm$0.05 & $<$0.16 & $<$4.7\\
\hline
\end{tabular}
\end{center}
\caption{-- \emph{Continue}.}
\normalsize
\end{table}

\end{document}